\documentclass[twocolumn,showpacs,floatfix,aps]{revtex4}
\usepackage{graphicx}
\usepackage{dcolumn}
\usepackage{bm}

\begin{document}
\title{Qubits with electrons on liquid helium}

\author{M.I. Dykman$^1$}
\email{dykman@pa.msu.edu},
\author{P.M. Platzman$^2$}
\author{P. Seddighrad$^1$}
\affiliation{$^1$Department of Physics and Astronomy,
Michigan State University, East Lansing, MI 48824\\
$^2$Bell
Laboratories, Lucent Technologies, Murray Hill, New Jersey 07974}
\date{\today}
\begin{abstract}
We study dissipation effects for electrons on the surface of liquid
helium, which may serve as qubits of a quantum computer. Each electron
is localized in a 3D potential well formed by the image potential in
helium and the potential from a submicron electrode submerged into
helium. We estimate parameters of the confining potential and characterize
the electron energy spectrum. Decay of the excited electron state
is due to two-ripplon scattering and to scattering by phonons in
helium. We identify mechanisms of coupling to phonons and estimate
contributions from different scattering mechanisms. Even in the
absence of a magnetic field we expect the decay rate to be $\alt
10^4$~s$^{-1}$. We also calculate the dephasing rate, which is due
primarily to ripplon scattering off an electron. This rate is $\alt
10^2$~s$^{-1}$ for typical operation temperatures.
\end{abstract}


\maketitle

\section{Introduction}
Much interest has attracted recently the idea of creating a
condensed-matter based quantum computer (QC). A major challenge is
to have a system that would have a sufficiently long relaxation
time and nevertheless could be controlled with high precision and
allow its quantum state to be measured. The proposed systems
include localized electron spins in semiconductor heterostructures
\cite{Loss-DV98,Imamoglu99,Si-Ge_spin00}, nuclear spins of
$^{31}$P donors \cite{Kane98} or $^{29}$Si nuclei
\cite{Yamamoto01} in a zero nuclear spin $^{28}$Si matrix,
electron states in a quantum dot excited by terahertz radiation
\cite{Sherwin99}, excitons in quantum dots
\cite{Steel_Sci00,Piermar01}, Josephson-junction based systems
\cite{Averin98,Tsai99,Lukens00,Orlando00,Han02,Urbina02},
electrons on helium surface \cite{PD99,DP01}, quantum dots coupled
via a linear support \cite{Lidar_supprt}, and trapped polar
molecules \cite{DeMille02}.

\begin{figure}
\includegraphics[width=2.4in]{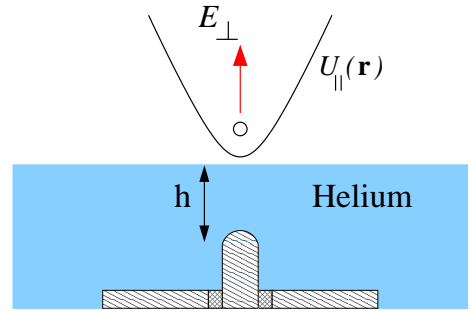}
\hfill
\caption{ A sketch of a micro-electrode submerged by the depth $h\sim
0.5\,\mu$m beneath the helium surface, with an electron localized
above it. The electron is driven by a field $E_{\perp}$ normal to the
surface. This field comes from the electrode and the parallel-plate
capacitor (only the lower plate of the capacitor is shown). The
in-plane electron potential $U_{_{\parallel}}({\bf r})$ is parabolic
near the minimum, with curvature determined by the electrode potential
(${\bf r}=(x,y)$ is the in-plane position vector). }
\label{fig:electrode}
\end{figure}

The system of electrons on the surface of superfluid $^4$He is
attractive, from the point of view of making a scalable quantum
computer, because (i) it has already been extensively studied
theoretically and experimentally \cite{Andrei_book}, (ii) the
electrons have extremely long relaxation time: they display the highest
mobility known in a condensed-matter system \cite{Shirahama-95}, and
(iii) the inter-electron distance is comparatively large, $\sim 1\,
\mu$m. To make a QC we suggested \cite{PD99,DP01} to
fabricate a system of micro-electrodes, which would be submerged beneath the
helium surface. Each electrode is supposed
to localize one electron above it, as seen in
Fig.~\ref{fig:electrode}, and to control this electron.

The two states of an electron qubit are the two lowest states of
quantized motion transverse to the surface. To further slow down the
already slow relaxation, we initially proposed to apply a magnetic
field $B_{\perp}$ normal to the surface. Then the estimated relaxation
time $T_2$ becomes as small as $10^{-4}$~s, for typical $B_{\perp}\sim
1.5$~T and temperatures $T\approx 10$~mK, whereas the clock frequency
of the computer $\Omega$ can be in the GHz range. This attracted attention of
experimentalists to the project
\cite{Lea_Fortschr,Goodkind01,Dahm_JLT02}.

In this paper we show that, even without a magnetic field, the
relaxation rate of a confined electron can be much less than that of a
free electrons. The dephasing rate can be even smaller than the
previous estimate for a strong magnetic field. This is due to large
level spacing in a 3D confining potential formed by a localizing
micro-electrode provided  the electrode is sufficiently thin. Electrodes of an
appropriate shape have already been fabricated \cite{Goodkind01}.
\hfill

For low temperatures, the major known dissipation mechanism is
scattering by surface capillary waves, ripplons. These waves are very
slow. Therefore a large distance between electron energy levels makes
it impossible to conserve energy and momentum in a one-ripplon decay
process. Decay of the excited electron state, i.e.  electron energy
relaxation may occur via scattering into two short-wavelength
ripplons.  We show that a very important role is played also by decay
processes, where the electron energy goes to phonons in helium. Such
phonons propagate nearly normal to the surface. We identify the major
mechanisms of electron-phonon coupling and analyze their contribution
to the decay rate. Dephasing is due primarily to scattering of
thermally excited ripplons off an electron. We find its temperature
dependence for different coupling mechanisms. We also investigate the
spectrum of sideband absorption in which a microwave-induced electron
transition is accompanied by creation or annihilation of a ripplon,
and analyze the related decrease of the intensity of the zero-ripplon
absorption line.

In Sec.~II below we analyze the energy spectrum of a confined electron
and discuss many-electron effects. In Sec.~III we discuss energy
relaxation rate for different mechanisms of electron-ripplon and
electron-phonon coupling. In Sec.~IV we consider dephasing
rate. Sec.~V deals with one-ripplon sidebands and the Debye-Waller
type factor in the zero-ripplon absorption line. In Sec.~VI we discuss
electron relaxation and dephasing from fluctuations in the underlying
electrodes. Sec.~VII contains concluding remarks.

\section{Electron states in one- and many-electron systems}
\subsection{Single-electron energy spectrum}

The quantum computer considered in this paper is based on a set of
electrons which reside in potential wells in free space above helium,
cf.  Fig.~\ref{fig:electrode}.
The electrons are prevented from penetrating into helium by a high
potential barrier $\sim 1$~eV at the helium surface.
For one electron, the potential well is formed by
the electrostatic image in helium, the potential from the electrode,
and also the potential created by the grounded plate and a parallel
plate above the electron layer (the latter is not shown in
Fig.~\ref{fig:electrode}).

We assume that the helium occupies the halfspace $z\leq 0$. The image
potential for an electron is $-\Lambda/z$, where $\Lambda=
(\varepsilon - 1)e^2/4(\varepsilon + 1)$, with $\varepsilon\approx
1.057$ being the dielectric constant of helium.  The energy spectrum
for 1D motion in such a potential is hydrogenic,
\begin{equation}
\label{hydrogenic_spectrum}
E_n=-R/n^2 \;(n=1,2,\ldots), \quad R=\Lambda^2m/2\hbar^2.
\end{equation}
The effective Rydberg energy $R$ is
$\approx 8$~K, and the effective Bohr radius is $r_B =
\hbar^2/\Lambda m \approx 76$~\AA ($m$ is the electron mass).

The electrode potential leads to Stark shift of the energy levels
(\ref{hydrogenic_spectrum}) \cite{Grimes-76} and to quantization
of motion parallel to the surface. A realistic estimate of this
potential and of the electron energy spectrum can be made by
modelling the electrode as a conducting sphere with a diameter
$2r_{\rm el}$ equal to the electrode diameter. The center of the
sphere is located at depth $h$ beneath the helium surface.
Typically we expect $h$ to be $\sim 0.5\,\mu$m, so that it largely
exceeds the distance from the electron to the surface $\sim r_B$.
For $z\ll h$ and for the in-plane distance from the electrode
$r\equiv (x^2+y^2)^{1/2}\ll h, (h^2-r_{\rm el}^2)^{1/2}$, the
electron potential energy is
\begin{equation}
\label{potential}
U({\bf r},z)\approx -{\Lambda\over z} + e{\cal
E}_{\perp}z + {1\over 2}m\omega_{_{\parallel}}^2r_{_{\parallel}}^2,
\end{equation}
with
\begin{eqnarray}
\label{frequency}
{\cal E}_{\perp}&=&V_{\rm el}r_{\rm el}h^{-2} + er_{\rm el}h(h^2-r_{\rm el}^2)^{-2},
\nonumber\\
\omega_{_{\parallel}}&=&(e{\cal E}_{\perp}/mh)^{1/2}.
\end{eqnarray}
Here, ${\bf r}=(x,y)$ is the electron in-plane position vector, and $V_{\rm
el}$ is the electrode potential. The second term in ${\cal E}_{\perp}$
comes from the image of the electron in the spherical electrode.

\begin{figure}
\includegraphics[width=2.8in]{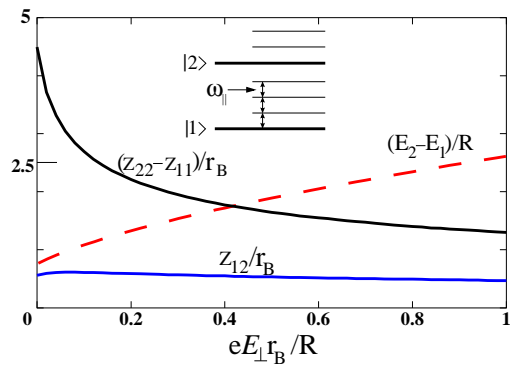}
\caption{Energy difference $E_2-E_1$ and matrix elements
$z_{nm}=\langle n|z|m\rangle$ of the electron coordinate normal to
helium surface on the wave functions of the ground and first
excited states of motion in the $z$-direction, $|1\rangle$ and
$|2\rangle$, vs. the overall pressing field $E_{\perp}$. The scaled field
$eE_{\perp}r_B/R=1$ for $E_{\perp}\approx 0.91$~kV/cm. The inset
shows the full energy level diagram.
Each level $E_n$ of $z$-motion gives rise to a set of energy
levels of vibrations parallel to helium surface, with typical
spacing $\hbar\omega_{_{\parallel}}$. } \label{fig:energies}
\end{figure}

In the approximation (\ref{potential}), the electron out-of-plane
and in-plane motions separate, with in-plane motion being just
harmonic oscillations. Variational calculations of the energy
spectrum of the out-of-plane motion were done earlier
\cite{Grimes-76}. The simple model (\ref{potential}) with an
infinite wall at $z=0$ describes the observed transition
frequencies with an error of only a few percent, which is
sufficient for the present purposes (more realistic models have
been discussed in literature, see
Refs.~\onlinecite{Cheng-Cole94,Nieto-00} and papers cited
therein). The full electron energy spectrum in the potential
(\ref{potential}) is sketched in the inset in
Fig.~\ref{fig:energies}. The two states of a qubit are the ground
and first excited states of motion transverse to the surface,
$|1\rangle$ and $|2\rangle$, both corresponding to the ground
state of in-plane vibrations.

In what follows, we characterize the electron state
$|i,\nu,m_{\nu}\rangle$ with the following 3 quantum numbers: $i=1,2$
enumerates the state of out-of-plane motion, $\nu=0,1,\ldots$ gives
the energy level of in-plane vibrations, and $m_{\nu}=0,1,\ldots,\nu$
enumerates degenerate vibrational states within this level.

\subsection{Choosing parameters of the many-electron system}

\subsubsection{Working frequency considerations}

For a multi-qubit multi-electrode QC, the depth $h$ by which the
controlling electrodes are submerged into helium, the
inter-electrode distances $d_{ij}$, and the electrode potentials
should be chosen in a way that would optimize performance of the
QC. This includes, in the first place, having a high working
frequency $\Omega_{\rm QC}$ and low relaxation rate $\Gamma$. The
frequency $\Omega_{\rm QC}$ is limited by the rate of single-qubit
operations and by the rate of excitation transfer between
neighboring qubits, which is determined by the qubit-qubit
interaction.

Single-qubit operations will be performed \cite{PD99} by applying
pulses of resonant microwave radiation, which cause transitions
between the states $|1\rangle$ and $|2\rangle$. The corresponding Rabi
frequency is $\Omega_R=e{\cal E}_m|z_{12}|/\hbar$, where ${\cal E}_m$
is the field amplitude. As seen from Fig.~\ref{fig:energies},
$|z_{12}|/r_B \agt 0.5$, and therefore even a comparatively weak field
${\cal E}_m=1$~V/cm gives $\Omega_R \agt 6\times 10^8$~s$^{-1}$.  This shows
that single-qubit operations should not limit $\Omega_{\rm QC}$ at
least at the level of $10^7-10^8$~Hz.

Because the wave functions of different electrons do not overlap,
the interaction between the qubits that we consider is dipolar, as
in liquid-state NMR quantum computers \cite{Chuang_book}. An
important feature of electrons on helium is that their
localization length normal to the surface $ r_B$ greatly exceeds
the atomic radius, which makes the dipole-dipole interaction
orders of magnitude stronger than the dipolar interaction in
atomic systems.

Of interest to us is the part of the qubit-qubit interaction that
depends on the states of the qubits. Two types of dipole moments
have to be distinguished. One is determined by the difference
$z_{11}- z_{22}$ of average distances of the electron from helium
surface in the states $|1\rangle$ and $|2\rangle$. The dipole
moment $e(z_{11}- z_{22})$ does not depend on time, if we take
into account time dependence of the wave functions, it can be
called ``static''. The interaction energy between the static
dipoles of the $i$th and $j$th qubits can be written as
$(1/4)U_{ij}^{\rm (st)}\sigma_z^i\,\sigma_z^j$, where
$\sigma_z^i=|2\rangle_i\,\langle 2|_i-|1\rangle_i\,\langle 1|_i$
is the operator of the difference of the state occupations for the
$i$th qubit, and
\begin{equation}
\label{static_dip}
U_{ij}^{\rm (st)} = e^2|z_{22}-z_{11}|^2/d_{ij}^3.
\end{equation}

The other dipole moment is associated with the $1\to 2$
transition. If we use time-dependent wave functions, it oscillates
in time at high frequency $\Omega_{12}=(E_2-E_1)/\hbar$. Resonant
interaction between such oscillating dipoles has energy
$(1/4)U_{ij}^{\rm (osc)}[\sigma_+^i\,\sigma_-^j+H.c.]$, where
$\sigma_+^i=[\sigma_-^i]^{\dagger}= 2|2\rangle_i\,\langle 1|_i$ is
the $1\to 2$ transition operator for the $i$th qubit, and
\begin{equation}
\label{osc_dip}
U_{ij}^{\rm (osc)} = e^2|z_{12}|^2/d_{ij}^3.
\end{equation}
The interaction between static and oscillating dipoles is nonresonant
and can be safely neglected.

The interactions (\ref{static_dip}) and (\ref{osc_dip}) allow
implementation of a CNOT two-qubit gate and of interqubit
excitation transfer, respectively \cite{PD99,DP01}. For a typical
dipole moment $er_B$, the interaction energy $e^2r_B^2/d_{ij}^3$
between the qubits separated by $d_{ij}=1\,\mu$m is $2\times
10^7$~Hz, in frequency units. This energy is very sensitive to
$d_{ij}$ and can be increased by reducing the inter-electron
distance. Eqs.~(\ref{static_dip}), (\ref{osc_dip}) apply for
$d_{ij}$ less than the distance from the electrons to the grounded
plate in Fig.~\ref{fig:electrode}; for larger $d_{ij}$ the
interaction is screened and falls down as $d_{ij}^{-5}$.  In
practice it means that the interqubit coupling is likely to be
limited to nearest and probably next nearest neighbors.

The matrix elements $z_{nm}$ depend on the overall field
$E_{\perp}$ that presses electrons against the helium surface.
They can be obtained by solving a one-dimensional Schr\"odinger
equation for the potential $-\Lambda z^{-1} +eE_{\perp}z$ with a
hard wall at $z=0$ [cf. Eq.~(\ref{potential}); we note that the
total field $E_{\perp}$ differs from the field ${\cal E}_{\perp}$
produced by one electrode, see below]. The results are shown in
Fig.~\ref{fig:energies}.

The difference $z_{22}-z_{11}$ sharply decreases with increasing field
for small $E_{\perp}$ because of field-induced squeezing of the wave
functions, which is particulalry strong for the wave function of the
excited state $|2\rangle$. The interplay between the squeezing and
better overlapping of the wave functions $|1\rangle$ and $|2\rangle$
with increasing field leads to a weak field dependence of $z_{12}$ for
$eE_{\perp}r_B/R\alt 1$. It is seen from Fig.~\ref{fig:energies} and
Eq.~(\ref{static_dip}) that, for weak pressing field $E_{\perp}<
300$~V/cm, the energy of the ``static'' interaction is higher than its
estimate given above by a factor varying from 20 to 4 with increasing
$E_{\perp}$, because of the large numerical value of
$(z_{22}-z_{11})/r_B$. It is also significantly higher than the energy
given by Eq.~(\ref{osc_dip}).

In a multi-qubit system, the fields ${\cal E}_{\perp}$ on different
electrodes are used to tune targeted qubits in resonance with
microwave radiation and with each other. In the simple case of one
microwave frequency, all these fields are nearly the same, they differ
by $\sim 1$~V/cm, or $\sim 1$\%. Therefore in Eqs.~(\ref{static_dip}),
(\ref{osc_dip}) we assumed that the matrix elements $z_{nm}$ are the
same for different qubits. Overall, for interelectron distances $d\alt
1\,\mu$m, the qubit-qubit interaction limits the clock frequency of
the quantum computer $\Omega_{\rm QC}$ to $10^7-10^8$~Hz.

\subsubsection{Limitations from many-electron effects}

The electron energy spectrum should be formed so as to minimize
the electron relaxation rate. One of the most ``dangerous''
relaxation processes is quasi-elastic scattering by capillary
waves on helium surface, ripplons, in which an electron makes a
transition between its states and a ripplon is emitted or
absorbed. This scattering is responsible for finite electron
lifetime $T_1$. Typical energies of appropriate ripplons are
extremely small, $\sim 10^{-3}$~K (see below). Therefore the
scattering can be eliminated for a one-qubit system, if none of
the excited vibrational levels of the state $|1\rangle$ is in
resonance with the ground vibrational level of the state
$|2\rangle$ shown with a bold line in Fig.~\ref{fig:energies}.

From Eq.~(\ref{frequency}), for a field ${\cal E}_{\perp} = 500$~V/cm
and $h=0.5\,\mu$m we have $\omega_{_{\parallel}}/2\pi\approx 2.1\times
10^{10}$~Hz $\approx 1.0$~K. Even though the spacing between
vibrational levels is less than the energy gap $E_2-E_1 \sim
6-10$~K, with so big $\omega_{_{\parallel}}$ it is easy to avoid
resonance between $E_2$ and an excited vibrational level of the state
1, i.e. between $E_2-E_1$ and  $n\hbar\omega_{_{\parallel}}$.

The situation becomes more complicated for a system of interacting
qubits.  The  interaction leads to coupling of in-plane vibrations
of different electrons. In a many-electron system the vibrational
energy spectrum becomes nearly continuous. One can think that each
vibrational level in Fig.~\ref{fig:energies} becomes a bottom of a
band of in-plane vibrational excitations. We will assume that the
width of the lowest band $\Delta_{_{\parallel}}$ is small compared
to $\omega_{_{\parallel}}$. The width of the $\nu$th band is then
$\sim \nu \Delta_{_{\parallel}}$ for not too large $\nu$.  To
avoid quasi-elastic scattering by ripplons, the electron energy
spectrum has to be discrete, i.e. vibrational bands should be well
separated from each other up to energies $E_2-E_1$, that is for
$\nu\sim (E_2-E_1)/\hbar\omega_{_{\parallel}}$. This means that
\begin{equation}
\label{bandwidth}
\Delta_{_{\parallel}} \ll \hbar\omega_{_{\parallel}}^2/(E_2-E_1).
\end{equation}

The value of $\Delta_{_{\parallel}}$ depends on the geometry of the
many-electron system. It can be found if the electrodes and the
electrons above them form a regular 2D array, or in other words, the
electrons form a Wigner crystal with the same lattice constant as the
electrodes. Then, if the phonon frequencies of the free-standing
crystal in the absence of the electrode potential are $\omega_{{\bf
k}j}$ (${\bf k}$ is the wave vector and $j=1,2$ is the branch number),
then the vibrational frequencies of the pinned crystal are
$(\omega_{{\bf k}j}^2+\omega_{_{\parallel}}^2)^{1/2}$. The phonon
bandwidth is small compared to $\omega_{_{\parallel}}$ provided
$\omega_{{\bf k}j}\ll \omega_{_{\parallel}}$, in which case
$\Delta_{_{\parallel}} = \max\omega_{{\bf k}j}^2/\omega_{_{\parallel}}
\sim \omega_p^2/\omega_{_{\parallel}}$, where $\omega_p=(2\pi
e^2n_e^{3/2}/m)^{1/2}$ is the characteristic zone-boundary frequency
of the free-standing Wigner crystal ($n_e$ is the electron density).

It follows from the above arguments and the condition
(\ref{bandwidth}) that quasi-elastic scattering will be eliminated
for a pinned Wigner crystal provided
\begin{equation}
\label{density_limit}
\omega_p^2\ll \hbar \omega_{_{\parallel}}^3/(E_2-E_1).
\end{equation}
This imposes an upper limit on the nearest neighbor spacing $d=\min
d_{ij}$, because $\omega_p\propto d^{-3/2}$. For a square lattice with
$d= 1\;\mu$m we have $\omega_p/2\pi\approx 6.3$~GHz.

\begin{figure}
\includegraphics[width=2.8in]{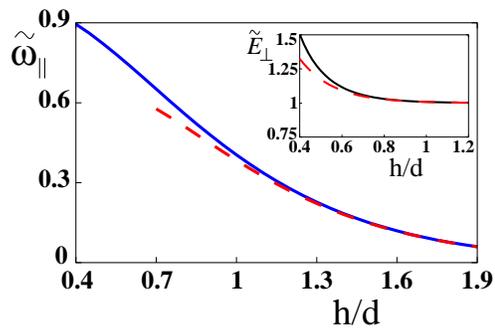}
\hfill

\vspace*{0.3cm}
\caption{In-plane frequency
$\tilde\omega_{_{\parallel}}=\omega_{_{\parallel}}/\omega_{_{\parallel}}'$
and normal to the surface field $\tilde E_{\perp}=
E_{\perp}/E_{\perp}'$ for an electron above a square array of
electrodes. The electron is localized at height $h$ above one of the
electrodes. The inter-electrode spacing is $d$. Electrodes are
modelled by small spheres, $r_{\rm el}/h\ll 1$, with same positive
potential $V_{\rm el}$. The scaling frequency
$\omega_{_{\parallel}}'=(eV_{\rm el}r_{\rm el}/mh^3)^{1/2}$ is given
by Eq.~(\protect\ref{frequency}) and corresponds to the limit $d\to
\infty$. The scaling field is $\tilde E_{\perp}'= 2\pi n_e V_{\rm
el}r_{\rm el}$. Asymptotic behavior of $\tilde\omega_{_{\parallel}}$
and $\tilde E_{\perp}$ for large $2\pi h/d$ is shown with dashed
lines.}
\label{fig:frequency}
\end{figure}

For the multi-electrode system, the frequency $\omega_{_{\parallel}}$
itself depends on the inter-electrode distance $d$. If the electrode
radius $r_{\rm el}$ is small compared to the depth $h$, the effect of
the electrostatic image in the electrode [in particular, the second
term in Eq.~(\ref{frequency}) for ${\cal E}_{\perp}$] can be
ignored. The overall potential of the electrode lattice at a distance
$z$ above helium surface ($z+h>0$) is
\begin{eqnarray}
\label{elcd_pot}
V({\bf r},z)=&&2\pi n_e V_{\rm el}r_{\rm
el}\sum^{\prime}\nolimits_{\bf G}G^{-1}\exp(i{\bf
Gr})e^{-G(z+h)}\nonumber\\
&&-2\pi n_e V_{\rm el}r_{\rm el}(z+h)
\end{eqnarray}
where ${\bf G}$ is the reciprocal lattice vector.

The dependence of $\omega_{_{\parallel}}$ on $h/d$ for a square
electrode array is shown in Fig.~\ref{fig:frequency} along with the
$z$-dependence of the total normal field from the electrodes.  The
electrostatic in-plane confinement is due to the spatial nonuniformity
of the electrode potential. Therefore $\omega_{_{\parallel}}$ falls
down as $2\pi (2eV_{\rm el}r_{\rm el}/md^3)^{1/2}\exp(-\pi h/d)$ for
large $2\pi h/d$. However, as seen from Fig.~\ref{fig:frequency},
$\omega_{_{\parallel}}$ remains close to the single-electrode value
(\ref{frequency}) for $h/d\alt 0.5$. This gives the desirable range of
the aspect ratio $h/d$.

The total perpendicular field on a localized electron $E_{\perp}$
comes from the electrodes and the capacitor that holds the system (its
lower plate is shown in Fig.~\ref{fig:electrode}). As we will see, the
field-induced squeezing of the electron wave functions (cf.
Fig.~\ref{fig:energies}) makes a significant effect on the electron
relaxation rate. Therefore $E_{\perp}$ should be minimized in order to
reduce relaxation effects. At the same time, the electrostatic
confinement (the frequency $\omega_{_{\parallel}}$) increases with the
increasing field from the electrodes. It would be good to compensate
the out-of-plane field $E_{\perp}$ while keeping the in-plane
potential as strongly confining as possible. This can be accomplished
using the field from the capacitor, which is uniform in the plane and
does not affect in-plane confinement.

The limitation on the compensating capacitor field comes from the
condition that the overall field behind the electron layer should
attract electrons to helium, otherwise they will leave the
surface. This field is formed not only by the externally applied
potentials, but also by the electron layer itself.  The total
averaged over ${\bf r}$ applied field in the electron plane should
therefore exceed $4\pi en_e$. In other words, the uniform
component $2\pi n_e V_{\rm el}r_{\rm el}$ of the electrode field
$-\partial_z V|_{z=0}$ (\ref{elcd_pot}) can be compensated down to
$4\pi en_e$. The remaining pressing field on the electron
$E_{\perp}$ becomes then $C\times 2\pi n_e V_{\rm el}r_{\rm el} +
4\pi e n_e$ with small $C$ ($C\approx 0.24$ for $h/d= 0.5$, as
seen from Fig.~\ref{fig:frequency}).

We note that the frequency $\omega_{_{\parallel}}$ can be further
increased electrostatically without increasing $E_{\perp}$ by using a
more sophisticated configuration of electrodes. Analysis of such
configurations is outside the scope of this paper. We note also that,
for sufficiently large $\omega_{_{\parallel}}$, the curvature of the
electrode potential (\ref{elcd_pot}) in the $z$-direction may become
substantial, particularly for highly excited states of out-of-plane
motion. However, for a typical $\omega_{_{\parallel}}/2\pi = 20$~GHz,
the effective curvature-induced change of the out-of-plane field for
lowest states $2m\omega_{_{\parallel}}^2r_B/e$ is only $\approx
14$~V/cm.


\subsection{Electrostatic force on helium}

Electric field from the electrodes and pressure from the electrons
(polaronic effect) lead to deformation of the helium surface. The
effect of the electrode potential can be easily estimated by
noticing that the dielectric constant of helium is close to one,
$\varepsilon - 1\approx 0.057 \ll 1$. Therefore if the surface is
raised by $\xi({\bf r})$, the associated change in the density
(per unit area) of the free energy of helium $\Delta {\cal F}$ in
the surface field ${\bf E}({\bf r})$ is $-(\varepsilon - 1)
E_{\perp}^2({\bf r})\xi({\bf r})/8\pi$. Bending of the surface is
counteracted by surface energy, with density $\sigma(\partial\xi/
\partial {\bf r})^2/2$, where $\sigma$ is the surface tension. The
competition between these two terms gives the height $\xi\sim
{\varepsilon - 1\over 8\pi} E_{\perp}^2 hd/\sigma$. For typical
$E_{\perp}= 3\times 10^2$~V/cm, $h=0.5\,\mu$m, and $d=1\,\mu$m
this gives a negligibly small $\xi < 10^{-10}$~cm. Therefore this
effect can be safely ignored.

\section{Decay of the excited electron state}
\subsection{The Hamiltonian of coupling to surface displacement}
The major mechanism of electron relaxation for low temperatures is
scattering by vibrations of the liquid helium surface. A complete
calculation of the energy of coupling to surface vibrations is
nontrivial.  The density profile of the interface between helium
and its vapor has a complicated form, with the 10\%/90\%
interfacial width $\approx 6-7$~\AA \, for low temperatures
\cite{Penanen-Pershan00}. As a consequence, even for a flat
surface the electron potential is more complicated than the simple
image potential $-\Lambda/z$ for $z>0$ and a sharp wall at $z=0$
(\ref{potential}) \cite{Cheng-Cole94}. In particular the repulsive
barrier is smooth, but it becomes high compared to the binding
energy $R$ already on the tail of the helium density distribution.
The spatial structure of surface excitations is complicated as
well. However, for excitations with sufficiently long wavelengths
to a good approximation the vibrating helium surface can still be
considered as a corrugated infinitely high potential wall. The
electron wave function is set equal to zero on the surface.

In this approximation the Hamiltonian $H_i$ of interaction of an
electron with surface vibrations is obtained by changing the electron
coordinates ${\bf r} \rightarrow {\bf r}, z\rightarrow z-\xi({\bf r})$
where $\xi({\bf r})$ is the surface displacement, see
Refs. \onlinecite{Platzman,Saitoh,Monarkha-Shikin82}. The interaction
is a series in the ratio $\xi/r_B$. Typically this ratio is very
small, $\sim 3\times 10^{-3}$ for thermal displacement with
characteristic wave numbers. Therefore to a good approximation $H_i$
can be expanded in $\xi$, keeping only lowest-order terms. The major
term, $H_i^{(1)}$, is linear in $\xi({\bf r})=\sum\nolimits_{\bf q}\xi_{\bf
q}e^{i{\bf qr}}$,

\begin{eqnarray}
\label{H_i1}
H_i^{(1)}=\sum_{\bf q} \xi_{\bf q}e^{i{\bf qr}}\,
\hat V_{\bf q},
\end{eqnarray}
with
\begin{eqnarray}
\label{V}
&&\hat V_{\bf q} =
-{i\over m}({\bf q}\cdot \hat{\bf p})\hat p_z -
{i\hbar\over 2m}q^2\hat p_z
+ eE_{\perp}
 +\Lambda q^2v_{\rm pol}(qz),\nonumber\\
&& v_{\rm pol}(x)=x^{-2}\left[1-xK_1(x)\right]
\end{eqnarray}
Here, $\hat {\bf p} = -i\hbar\partial_{\bf r}$ is the 2D electron
momentum, and $\hat p_z=-i\hbar\partial_z$. The first two terms in the
operator $\hat V_{\bf q}$ describe a {\it kinematic} interaction,
which arises because the electron wave function is set equal to zero
on a non-flat surface.  The term $v_{\rm pol}(qz)$ describes the change
of the polarization energy due to surface curvature
\cite{Saitoh,Monarkha-Shikin82} ($K_1(x)$ is the modified Bessel
function).

The quadratic in $\xi$ coupling is
\begin{equation}
\label{H_i2}
H_i^{(2)}=\sum\nolimits_{{\bf q}_1,{\bf q}_2} \xi_{{\bf q}_1}\xi_{{\bf q}_2}
\exp[i({\bf q}_1+{\bf q}_2){\bf r}]\, \hat V_{{\bf q}_1{\bf q}_2}.
\end{equation}
As in the case of linear coupling, it also has kinematic
\cite{Monarkha-Shikin82} and polarization parts,
\begin{eqnarray}
\label{V_2}
\hat V_{{\bf q}_1{\bf q}_2}=\hat V_{{\bf q}_1{\bf q}_2}^{\rm (k)}+
\hat V_{{\bf q}_1{\bf q}_2}^{\rm (pol)},\; \hat V_{{\bf q}_1{\bf
q}_2}^{\rm (k)}=-({\bf q}_1{\bf q}_2)\,p_z^2/2m,
\end{eqnarray}
with
\begin{eqnarray}
\label{V_2pol}
\hat V_{{\bf q}_1{\bf q}_2}^{\rm (pol)}&&=-\Lambda
z^{-3}[1-u(q_1z)-u(q_2z)+u(|{\bf q}_1+{\bf q}_2|z)],\nonumber\\
&&u(x)=x^2K_2(x)/2
\end{eqnarray}
\subsubsection{Coupling to ripplons}

The biggest contribution to surface vibrations comes from
capillary waves, ripplons.  The displacement $\xi_{\bf q}$ is related
to the creation and annihilation operators of ripplons by
\[\xi_{\bf q}=S^{-1/2}(\hbar q/2\rho\omega_q)^{1/2}
(b_{\bf q}+b^{\dagger}_{-\bf q}),\]
where $S$ is the area of the system, $\rho$ is the helium density, and
the ripplon frequency $\omega_q=(\sigma q^3/\rho)^{1/2}$ for $q \gg
(\rho g/\sigma)^{1/2}$.

The change of variables used to take into account the hard wall
potential on helium surface leads also to extra terms in the kinetic
energy of ripplons coupled to the electron, which is yet another
source of electron-ripplon coupling \cite{Monarkha-Shikin82}. Compared
to similar terms in Eqs.~(\ref{H_i1}), (\ref{H_i2}), these terms have an
extra parameter $\omega_q m/\hbar q^2$, which is extremely small for
typical $q$.

There are several limitations on the wave numbers $q$ of ripplons for
which the electron-ripplon coupling has the form (\ref{H_i1}),
(\ref{H_i2}). Monarkha and Shikin \cite{Monarkha-Shikin82} argue that
essentially $qr_B$ should be $\lesssim 1$. Clearly, $q$ should be
small compared to the reciprocal width of the helium liquid-vapor
interface and the reciprocal decay length of the electron wave
function into helium (note that there is no factor $2\pi$ here,
because a capillary wave with wave number ${\bf q}$ decays into helium
as $\exp(qz)$, for a sharp interface). Both lengths are of order of a
few angstroms, which means that the large-$q$ cutoff $q_{\max}$ should be
$\lesssim 10^7$~cm$^{-1}$.

A cutoff at $10^7$~cm$^{-1}$ is consistent also with the condition
that $H_i^{(2)}$ (\ref{H_i2}) be small. To first order in the
kinematic part of $H_i^{(2)}$, which dominates for large $q$, the
relative change of the electron kinetic energy for motion transverse
to the surface for $T=0$ is
\[\delta K/K =\hbar q_{\max}^{7/2}/14\pi (\sigma\rho)^{1/2}.\]
This gives $\delta K/K\approx 3\times 10^{-4}$ for $q_{\max}=
10^7$~cm$^{-1}$ (for $q_{\max}=10^8$~cm$^{-1}$ the correction would be
equal to 1). Presumably the interaction with ripplons as a whole and
in particular inelastic scattering by ripplons falls down for $q$ much
bigger than $10^7$~cm$^{-1}$, because for such momentum transfer an
electron ``resolves'' atomic structure of helium. Finding $H_i$ for
such $q$ requires a full calculation of the ripplon-induced modulation
of the electron potential for the diffuse helium surface, which is not
the subject of the present paper.

In what follows we will use spectroscopic notations and evaluate the
decay rate as $\Gamma = 1/2T_1$, where $T_1$ is the lifetime of the
excited state. Defined in this way, $\Gamma$ gives the decay rate of the
off-diagonal matrix element $\rho_{12}$ of the electron density
matrix and the decay-induced broadening of the absorption line.

\subsection{One-ripplon decay}
The important consequence of strong in-plane confinement is that
it essentially eliminates decay processes in which an electron
transition is accompanied by emission or absorption of a ripplon.
This happens because ripplons are very slow, and energy
conservation in a transition requires transfer of too large
momentum for an electron to accommodate.

For low temperatures, $k_BT\ll
\hbar\omega_{_{\parallel}}$, qubit relaxation is due to electron
transitions $|2,0,0\rangle \to |1,\nu,m_{\nu}\rangle$ from the ground
vibrational level of the state $|2\rangle$ of $z$-motion into excited
states of in-plane vibrations in the state $|1\rangle$ of $z$-motion,
see Fig.~\ref{fig:energies}.
Minimal energy transfer is of the order of the vibrational level
spacing $\hbar\omega_{_{\parallel}}$. It corresponds to a transition
into vibrational states with the energy closest to $E_2$ from below,
i.e. with $\nu=\nu_c \equiv {\rm int}[(E_2-E_1)/\hbar
\omega_{_{\parallel}}]$. The squared matrix element of the transition
$|\langle \nu=m_{\nu}=0|\exp(i{\bf qr})|\nu_c,m_{\nu_c}\rangle|^2$ is
exponentially small for $q\gg \nu_c^{1/2}/a_{_{\parallel}}$, where
\begin{equation}
\label{aparallel}
a_{_{\parallel}}=(\hbar/m\omega_{_{\parallel}})^{1/2}
\end{equation}
is the electron  in-plane localization length.

The frequency of ripplons with $q=\nu_c^{1/2}/a_{_{\parallel}}$ is
much less than $\omega_{_{\parallel}}$ provided
\[
\omega_{_{\parallel}}\gg
(\sigma/\rho)^{1/2}[m(E_2-E_1)/\hbar^2]^{3/4}.\]
This inequality is
satisfied already for $\omega_{_{\parallel}}/2\pi \agt$ 0.2~GHz,
whereas a typical $\omega_{_{\parallel}}$ for confined electrons is
$ \sim 20$~GHz. Therefore one-ripplon decay is exponentially
improbable. This result does not change for a many-electron system
provided the bands of in-plane vibrations are narrow, as discussed in
Sec.~II~B, see Eq.~(\ref{bandwidth}).

\subsection{Two-ripplon decay}

Even for large separation between electron energy levels, where
one-ripplon decay processes are exponentially suppressed by the
restriction on the transferred momentum, decay into two ripplons may
still be possible \cite{Dykman78,Monarkha78}. Indeed, each of the wave
vectors ${\bf q}_1,{\bf q}_2$ of the participating ripplons can be
big, this is only their sum that is limited by the typical reciprocal
electron wavelength. The energy and momentum conservation law require
than that the ripplons have nearly same frequencies and propagate in
opposite directions, ${\bf q}_1\approx -{\bf q}_2$ and
$\omega_{q_1}\approx\omega_{q_2} \approx \delta E/2\hbar$, where
$\delta E$
is the change of the electron energy.

A typical minimal value of $\delta E$ for the decay of the electron
state $|2\rangle$ is determined by the distance between the energy
levels of in-plane vibrations $\hbar\omega_{_{\parallel}}$.  The
ripplon frequency $\omega_q/2\pi$ becomes equal to typical
$\omega_{_{\parallel}}/4\pi=10$~GHz for $q=1.2\times 10^7$~cm$^{-1}$,
i.e. presumably beyond the range of applicability of the
electron-ripplon coupling theory (\ref{H_i1}), (\ref{H_i2}). Therefore
we use this theory to estimate the rate of decay only with minimal
energy transfer, i.e. we are interested in ripplon-induced transitions
$|2,0,0\rangle \to |1,\nu,m_{\nu}\rangle$ with $\nu=\nu_c$. In this case
\[\delta E=E_2-E_1-\nu_c\hbar \omega_{_{\parallel}}.\]


In calculating the total rate of decay into different states of the $\nu$th
vibrational energy level of a two-dimensional oscillator it is
convenient to use the relation
\begin{eqnarray}
\label{g_nu} g(\nu,q)=&&\sum\nolimits_{m_{\nu}}\vert\langle j
,0,0|e^{i{\bf
qr}}|j,\nu,m_{\nu}\rangle\vert^2\nonumber\\
&&=x^{\nu}e^{-x}/\nu!,\quad x=q^2a_{_{\parallel}}^2/2
\end{eqnarray}
independent of $j=1,2$.

\begin{figure}
\includegraphics[width=2.8in]{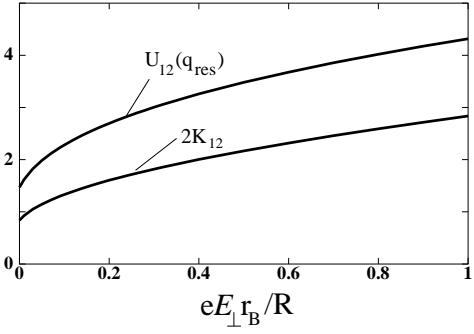}
%
%
%
\caption{Scaling factors $K_{12}$ (\protect\ref{two-width}) and
$U_{12}(q)$ (\protect\ref{pol-2rippl}) in the estimated probabilities
of scattering into two ripplons due to the kinematic and polarization
couplings, respectively. The ripplons propagate in opposite directions
with nearly same wave numbers $q_{\rm res}$ given by the energy
conservation condition $2\omega_q=\delta E/\hbar$ for $q=q_{\rm
res}$. Only transitions with smallest energy transfer $\delta E$ are
taken into account and the approximation of an infinite sharp potential wall
for an electron on helium surface is used. The data for $U_{12}$ refer
to $q_{\rm res}=3.5/r_B$, which corresponds to $\delta E/2\pi\hbar
\approx 5$~GHz.}
\label{fig:2ripplon}
\end{figure}

For the kinematic two-ripplon coupling, which is given by Eq.~(\ref{H_i2}) with
$\hat V_{{\bf q}_1{\bf q}_2}=\hat V_{{\bf q}_1{\bf q}_2}^{\rm (k)}$,
the decay rate $\Gamma_{2r}^{\rm (k)}$ with account taken of
(\ref{g_nu}) is
\begin{equation}
\label{two-width}
\Gamma_{2r}^{\rm (k)} = {K_{12}^2R^2q_{\rm res}^{7/2}\over 24\pi
a_{_{\parallel}}^2\rho^{1/2}\sigma^{3/2}},\; K_{12}={\langle
1|p_z^2/2m|2\rangle \over R}.
\end{equation}
Here, $q_{\rm res}$ is the ripplon wave vector given by the energy
conservation law, $\omega_{q_{\rm res}}=\delta E/2\hbar$, and we
assumed that $q_{\rm res}\gg \nu_c^{1/2}/a_{_{\parallel}}$, which is the
condition for the ripplons created in the transition to propagate in
opposite directions. The scaled matrix element of the kinetic energy
$K_{12}$ is shown in Fig.~\ref{fig:2ripplon}.

The rate $\Gamma_{2r}^{\rm (k)}$ depends on $q_{\rm res}$ and
therefore on $\delta E$ very steeply, $\Gamma_{2r}^{\rm (k)}\propto
\delta E^{7/3}$. For $\delta E=\hbar\omega_{_{\parallel}}/4$ and
$\omega_{_{\parallel}}/2\pi = 20$~GHz we have $q_{\rm res}\approx
4.6\times 10^6$~cm$^{-1}$ and $\Gamma_{2r}^{\rm (k)} = 7.6\times 10^2-
3.8\times 10^3$~s$^{-1}$ for the pressing field
$E_{\perp}=0-300$~V/cm. This value can be decreased by reducing
$\delta E$. However, Eq.~(\ref{two-width}) is probably an overestimate
even for the $\delta E$ used above, because it is based on the
approximation of an infinite-wall potential for an electron and the
assumptions that the helium surface is sharp.

The expression for the two-ripplon decay rate $\Gamma_{2r}^{\rm
(pol)}$ due to the polarization two-ripplon interaction (\ref{V_2pol})
has the same form as Eq.~(\ref{two-width}). Just the factor
$K_{12}^2q_{\rm res}^{7/2}$ in Eq.~(\ref{two-width}) has to be
replaced with $U_{12}^2(q_{\rm res})/r_B^4q_{\rm res}^{1/2}$, where
$U_{12}$ is determined by the matrix element of $2r_B^3\hat V_{{\bf q}_1{\bf
q}_2}^{\rm (pol)}/\Lambda$ (\ref{V_2pol}) on the functions
$|1\rangle, |2\rangle$. The major contribution to this matrix element
comes from the range of comparatively small $z$, and therefore for an
estimate one can replace $K_2(|{\bf q}_1+{\bf q}_2|z)$ with its
small-$z$ limit in (\ref{V_2pol}). Then
\begin{equation}
\label{pol-2rippl}
U_{12}(q)= 2r_B^3\langle 1|z^{-3}[2-q^2z^2K_2(qz)]|2\rangle.
\end{equation}

The coefficient $U_{12}$ as given by Eq.~(\ref{pol-2rippl}) is shown
in Fig.~\ref{fig:2ripplon}. For $\delta E$ and $\omega_{_{\parallel}}$
chosen above we have $\Gamma_{2r}^{\rm (pol)}/\Gamma_{2r}^{\rm (k)}
\sim 0.1$ for $E_{\perp}= 0- 300$~V/cm. The rate
$\Gamma_{2r}^{\rm (pol)}$ grows much slower then $\Gamma_{2r}^{\rm
(k)}$ with increasing $q_{\rm res}$ (and thus with increasing $\delta
E$).

Besides the terms $\Gamma_{2r}^{\rm (k)}$ and $\Gamma_{2r}^{\rm
(pol)}$ due to purely kinematic and polarization mechanisms, there is
a contribution to the decay rate from the interference of these two
mechanisms. It is smaller then $\Gamma_{2r}^{\rm (k)}
+\Gamma_{2r}^{\rm (pol)}$ and will not be discussed.

\subsection{Phonon-induced decay}

An important channel of inelastic electron scattering is decay into
phonons in helium. For a typical energy transfer $\delta E
\sim\hbar\omega_{_{\parallel}}$, the wave numbers of the phonons
participating in the decay are $\sim \omega_{_{\parallel}}/v_s$ ($v_s$
is the sound velocity in helium). On the other hand, the in-plane
momentum transfer is limited to $\sim \hbar/a_{_{\parallel}} \ll
\hbar\omega_{_{\parallel}}/v_s$.  As a result, only phonons
propagating nearly normal to the surface (in the $z$-direction) may be
excited in a one-phonon decay (cf. Ref.~\onlinecite{Dykman78}).

\subsubsection{Decay due to phonon-induced surface displacement}
Two coupling mechanisms are important for decay into phonons. One
is related to phonon-induced displacement of the helium surface.
This mechanism can be quantitatively described in the
approximation of a sharp helium boundary, which provides an
infinitely high potential barrier for electrons. The coupling is
given by Eqs.~(\ref{H_i1}), (\ref{V}) with $\xi_{\bf q}$ being a
phonon-induced component of the surface displacement.
As in the case of coupling to ripplons, it would be
unreasonable to use it for short-wavelength phonons, in particular
for phonons with $q_z\gg 10^7$~cm$^{-1}$. For typical
$\omega_{_{\parallel}}/2\pi = 20$~GHz we have $q_z\sim
\omega_{_{\parallel}}/v_s\sim 5\times 10^6$~cm$^{-1}$. Therefore
we will again consider decay of the state $|2,0,0\rangle$ into
closest lower-energy states $|1,\nu,m_{\nu}\rangle$.

For typical ${\bf q}\equiv (q_x,q_y)\sim 1/a_{_{\parallel}}$ and
$q_z\alt \omega_{_{\parallel}}/v_s$ we have $\sigma q^2/\rho v_s^2q_z
\ll 1$. This inequality allows one to think of helium surface as a
free boundary for phonons and to ignore coupling between phonons and
ripplons \cite{Edwards74,Lastri95}. Then surface displacement is
simply related to the Fourier components $u_{\bf Q}$ of the phonon
displacement field [here, $ {\bf Q}=({\bf q},q_z)$ is the 3D phonon
wave vector, and $u_{\bf Q}$ is the displacement along ${\bf Q}$].  In
turn, $u_{\bf Q}$ is related to the operators of creation and
annihilation of phonons in a standard way,
\begin{equation}
\label{phonon_coordinates}
u_{\bf Q} = (\hbar/2\rho V v_s Q)^{1/2}(c_{\bf Q} - c^{\dagger}_{-\bf Q})
\end{equation}
($V$ is the volume of helium).

With (\ref{phonon_coordinates}), we obtain the rate of decay
$|2,0,0\rangle \to |1,\nu_c,m_{\nu}\rangle$ due to phonon-induced
surface displacement $\Gamma^{\rm (s)}_{\rm ph}$ in the form
\begin{eqnarray}
\label{phonon1}
\Gamma^{\rm (s)}_{\rm ph}=&& \left(8\pi^2\rho v_s\delta E\right)^{-1}\nonumber\\
&&\times\sum_{m_{\nu}=0}^{\nu_c}\int d{\bf q}\, \vert\langle2,0,0|e^{i{\bf
qr}}\hat V_{\bf q}|1,\nu_c,m_{\nu}\rangle\vert^2.
\end{eqnarray}

We start with the contribution to $\Gamma^{\rm (s)}_{\rm ph}$ from the
kinematic  terms in $\hat V_{\bf q}$ [the first two terms in
Eq.~(\ref{V})].  Taking into account that the diagonal with
respect to out-of-plane motion matrix element 
\begin{eqnarray*}
&&\langle
j,0,0|\exp(i{\bf qr})[({\bf q}\hat{\bf p})+\hbar
q^2/2]|j,\nu,m_{\nu}\rangle\\&&= -\nu m\omega_{_{\parallel}}\langle
j,0,0|\exp(i{\bf qr})|j,\nu,m_{\nu} \rangle \quad (j=1,2)
\end{eqnarray*}
and using Eq.~(\ref{g_nu}), we obtain for the kinematic contribution
\begin{eqnarray}
\label{phonon-kinet}
\Gamma^{\rm (s;k)}_{\rm ph}\approx
(E_2-E_1)^2z_{12}^2\,{\nu_c^2 m^3\omega_{_{\parallel}}^3\over 4\pi\rho v_s
\hbar^3\delta E}.
\end{eqnarray}
The numerical value of $\Gamma^{\rm (s;k)}_{\rm ph}$ is $7.8\times
10^2$~s$^{-1}$ for $E_{\perp}=0$,
$\omega_{_{\parallel}}/2\pi=21.1$~GHz, and $\delta E\approx
\hbar\omega_{_{\parallel}}$ ($\nu_c=5$ in this case). It goes up to
$\sim 1.5\times 10^4$~s$^{-1}$ for $E_{\perp}= 300$~V/cm and
$\omega_{_{\parallel}}/2\pi=20.6$~GHz (in this case $\nu_c = 12$). The
values of $\omega_{_{\parallel}}$ were adjusted here to meet the
condition $\delta E=E_2-E_1-\nu_c\hbar\omega_{_{\parallel}}\approx
\hbar\omega_{_{\parallel}}$ for the energy spectrum calculated for a
sharp helium boundary; the real level spacing is a few percent smaller
\cite{Grimes-76,Cheng-Cole94,Nieto-00}, leading to a slightly smaller
$\Gamma^{\rm (s;k)}_{\rm ph}$ for $\omega_{_{\parallel}}/2\pi \sim
20$~GHz. We expect a more significant change (reduction) of
$\Gamma^{\rm (s;k)}_{\rm ph}$ due to diffuseness of helium surface.

The contribution to $\Gamma^{\rm (s)}_{\rm ph}$ from the polarization term
in $\hat V_{\bf q}$ [the last term in Eq.~(\ref{V})] has the form
\begin{eqnarray}
\label{phonon-polar} \Gamma^{\rm (s; pol)}_{\rm ph}\approx {4R^2r_B^2\over
\nu_c!\,\pi\rho v_s\, \delta E\,a_{_{\parallel}}^6}\int_0^{\infty}
dx\,e^{-x}x^{\nu_c+2}v^2(x),
\end{eqnarray}
where $v^2(x)=|\langle 1|v_{\rm
pol}[(2x)^{1/2}z/a_{_{\parallel}}]|2\rangle|^2$. The numerical value
of $\Gamma^{\rm (s; pol)}_{\rm ph}$ is $\sim 7\times 10^2$~s$^{-1}$
for $E_{\perp}=0$ and goes up to $\sim 7\times 10^3$~s$^{-1}$ for
$E_{\perp}= 300$~V/cm (we used same $\omega_{_{\parallel}}$ as in the
above estimate of $\Gamma^{\rm (s;k)}_{\rm ph}$).

There exists also a contribution to $\Gamma^{\rm (s)}_{\rm ph}$
(\ref{phonon1}) from the interference term, which is bilinear in the
polarization and kinematic interaction energies (\ref{V}) (the terms
$\Gamma^{\rm (s;k,pol)}_{\rm ph}$ are quadratic in these interactions). It
can be obtained from (\ref{phonon1}) in the same way as
$\Gamma^{\rm (s; k,pol)}_{\rm ph}$. It is positive and of the same order of
magnitude as $\Gamma^{\rm (s; k,pol)}_{\rm ph}$.

The scattering rate $\Gamma^{\rm (s)}_{\rm ph}$ can be reduced by
decreasing the pressing field $E_{\perp}$. More importantly, it
can also be reduced by increasing the frequency
$\omega_{_{\parallel}}$, in which case the wavelength of the
phonons to which the energy is transferred will become smaller
then the width of the diffusive layer on helium surface (in fact,
the above calculation probably already overestimates the
scattering rate). A transition to such frequency can be
accomplished with a magnetic field applied transverse to the
surface, as initially suggested in Ref.~\onlinecite{PD99}.

\subsubsection{Decay due to phonon-induced modulation of the helium
dielectric constant}

Another mechanism of coupling to phonons is through phonon-induced
modulation of the image potential of an electron. It results from the
modulation of the helium density $\delta \rho$ and related modulation
of the dielectric constant $\delta\varepsilon$. It is reasonable to
assume that, for long wavelength phonons, $\delta \varepsilon
=(\varepsilon -1)\delta \rho/\rho$. To lowest order in $\varepsilon
-1, \delta\varepsilon$ the coupling energy is
\[H_i^{\rm (d)}= -{1\over 8\pi}\int d{\bf R}'\,
\delta\varepsilon({\bf R}')\,E^2({\bf R}'; {\bf R}).\;\]
Here the integration goes over the space occupied by helium, ${\bf
R}\equiv ({\bf r},z)$ is the 3D position vector, and ${\bf E}({\bf
R}';{\bf R})$ is the electric field at ${\bf R}'$ created by an
electron located at a point ${\bf R}$.

The coupling Hamiltonian can be written in the form
\begin{eqnarray}
\label{dielectric_coupling}
H_i^{\rm (d)}=\sum\nolimits_{\bf Q}u_{\bf Q}\exp(i{\bf qr})
\hat V^{\rm (d)}_{\bf Q},\;
\hat V^{\rm (d)}_{\bf Q}=i\Lambda\,
q\,Qv^{\rm (d)}
\end{eqnarray}
with $v^{\rm (d)}\equiv v^{\rm (d)}(q,q_z,z)$ being
\begin{eqnarray}
\label{v^d}
v^{\rm (d)}=\int\nolimits_0^{\infty}dz'(z+z')^{-1}e^{-iq_zz'}
K_1\left[q(z+z')\right].
\end{eqnarray}

As in the case discussed in the previous section, the coupling
(\ref{dielectric_coupling}) gives rise to phonon-induced electron
transitions between electron energy levels accompanied by emission of
a phonon. Here, too, typical in-plane wave numbers of emitted
phonons $q$ are much less than the normal to the surface wave number
$q_z\approx \delta E/\hbar u$. The expression for the corresponding
decay rate $\Gamma^{\rm (d)}_{\rm ph}$ has the form
\begin{eqnarray}
\label{Gamma_diel}
\Gamma_{\rm ph}^{\rm (d)}\approx {R^2\,\delta E\,r_B^2\over
\pi\hbar^2\rho v_s^3}\,\int\nolimits_0^{\infty}dq\,q^3|\langle
2|v^{\rm (d)}|1\rangle|^2g(\nu_c,q)
\end{eqnarray}
Evaluation of the integral is largely simplified by the fact that the
function $q^3g(\nu,q)$ sharply peaks at $q=q_{\nu}\approx (2\nu
+3)^{1/2}/a_{_{\parallel}}$. Therefore, with an error less than 10\%
one can replace $v^{\rm (d)}$ in (\ref{Gamma_diel}) by its value
(\ref{v^d}) for $q=q_{\nu_c}$

For $\omega_{_{\parallel}}/2\pi = 20$GHz and $\delta
E=\hbar\omega_{_{\parallel}}$, the value of $\Gamma_{\rm ph}^{\rm
(d)}$ varies from $\sim 1\times 10^4$~s$^{-1}$ to $\sim 6\times
10^4$~s$^{-1}$ with $E_{\perp}$ increasing from 0 to 300
V/cm. However, these values have to be taken with care. The integrand
in $v^{\rm (d)}$ (\ref{v^d}) is a fast oscillating function of $z'$ on
the characteristic scale $z'\sim r_B$, because typically $q_zr_B \gg
1$ ($q_zr_B\approx 4$ for chosen $\omega_{_{\parallel}}$). In
addition, the matrix element of $v^{\rm (d)}$ in (\ref{Gamma_diel})
has an integrable singularity for $z=z'=0$ (the wave functions
$\psi_n(z)\propto z$ for $z\to 0$).  As a result, a significant
contribution to the matrix element comes from small distances from the
helium surface, $z'\ll r_B$.  Changing, in view of diffuseness of
helium surface, the limit of integration in (\ref{v^d}) from $z'=0$ to
a more reasonable $z'=r_B/10$ reduces the value of $\Gamma_{\rm
ph}^{\rm (d)}$ by a factor of 3.

The decay rate $\Gamma_{\rm ph}^{\rm (d)}$ decreases with the
increasing $\delta E$ roughly as $1/\delta E$ (and even faster, in
view of the ``dead'' layer on the diffuse surface). For higher
$\delta E$ and, respectively, for higher wave numbers of resonant
phonons, the simple approximation (\ref{dielectric_coupling}) no
longer describes the electron-phonon interaction. Therefore, as in
the case of scattering due to phonon-induced surface deformation,
a way to reduce the scattering rate is to increase the frequency
of in-plane vibrations.

Full coupling to phonons is given by the sum of all couplings
discussed in this section, with $\hat V_{\bf q}$ in
Eq.~(\ref{phonon1}) replaced by $\hat V_{\bf q}+ \alpha_{\bf Q}\hat
V_{\bf Q}^{\rm (d)}$ with $\alpha_{\bf Q}\approx 1$ for typical ${\bf
Q}$. The total rate of scattering by phonons contains cross-terms
which describe interference of different coupling mechanisms. As
mentioned above, we omit these terms, because they do not change the
overall estimate of the rate. We note that an interesting situation
may occur if one of the transition frequencies of the electron comes
in resonance with the roton energy. In this case we expect an increase
of the decay rate. Observing it would be a direct demonstration of
coupling to volume excitations in helium.

\section{Dephasing due to ripplon scattering}

In addition to depopulation of the excited state of a qubit, electron
coupling to excitations in liquid helium leads also to decoherence or
dephasing, i.e. to decay of the phase difference between the qubit
states $|2,0,0\rangle$ and $|1,0,0\rangle$. The mechanism of this
decay is random modulation, by thermal fluctuations in helium, of the
distance between the energy levels 1 and 2. In other terms it can be
described as quasi-elastic scattering of thermal excitations off an
electron. The scattering is different in different electron
states. Therefore it randomizes the phase difference between the wave
functions of the states without causing interstate transitions. The
corresponding mechanism is known for defects in solids
\cite{Stoneham75} as modulational or Raman broadening. For electrons
on helium it was discussed in Refs. \onlinecite{PD99,Dykman78}.

Dephasing comes primarily from coupling to ripplons, because these
excitations are soft. A typical wave number $q_r$ and frequency
$\omega_r$ of ripplons coupled to an electron are given by
\begin{equation}
\label{omega_r}
q_r=1/a_{_{\parallel}},\; \omega_r\equiv
\omega_{q_r} = (\sigma/\rho)^{1/2}q_r^{3/2}.
\end{equation}
For $\omega_{_{\parallel}}/2\pi =20$~GHz we have $\omega_r/2\pi
\approx 4.8\times 10^7$~Hz $\approx 2.3$~mK.
Therefore even for temperatures as low as 10 mK ripplon occupation
numbers are large. To the lowest order of perturbation theory,
quasi-elastic ripplon scattering by an electron is determined by
two-ripplon coupling, with the Hamiltonian
\begin{equation}
\label{H_iqe}
H_i^{\rm (qe)}=\sum_{j=1,2}\sum_{\bf q,q'}v_{{\bf qq'}j}b_{\bf
q}^{\dagger}b_{\bf q'}|j,0,0\rangle\langle j,0,0|.
\end{equation}
Individual terms in the sum over ${\bf q,q}'$ describe scattering of a
ripplon with wave number ${\bf q}'$ into a ripplon with wave
number ${\bf q}$. The momentum is transferred to the electron, and no
transitions between electron states occur. We will consider terms with
${\bf q}\neq {\bf q}'$; the term with ${\bf q}={\bf q}'$ gives the
shift of the electron energy levels.

The matrix elements $v_{{\bf qq'}j}$ are linear in the parameters of
the direct two-ripplon coupling $H_i^{(2)}$ (\ref{H_i2}) and quadratic
in the parameters of the one-ripplon coupling $H_i^{(1)}$ (\ref{H_i1}),
\begin{widetext}
\begin{eqnarray}
\label{qe_coupling}
v_{{\bf qq'}j}\approx{\hbar (qq')^{1/2}\over
S\rho(\omega_q\omega_{q'})^{1/2}}\left[\langle j,0,0|\hat V_{-{\bf
q},{\bf q}'}|j,0,0\rangle e^{-({\bf q}-{\bf
q}')^2a_{_{\parallel}}^2/4}
-\sum\nolimits_{\nu,m_{\nu}}\left( {\cal V}^{j\nu m_{\nu}}_{{\bf q\,q}'} +
{\cal V}^{j\nu m_{\nu}}_{{\bf -q}'\,{\bf -q}}\right) 
(\hbar\nu\omega_{_{\parallel}})^{-1}\right]
\end{eqnarray}
\end{widetext}
where $
{\cal V}^{j\nu m_{\nu}}_{{\bf q\,q}'}=V_{{\bf q}}^{j\nu
m_{\nu}}\left(V_{{\bf q}'}^{j\nu
m_{\nu}}\right)^*$ and 
$V_{{\bf q}}^{j\nu m_{\nu}}= \langle j,0,0|\hat V_{-{\bf
q}}e^{-i{\bf qr}}|j,\nu,m_{\nu}\rangle$.
In calculating
renormalization of the parameters $v_{{\bf qq'}j}$ due to one-ripplon
coupling we disregarded the contribution from virtual transitions into
different states of out-of-plane motion $|j'\rangle$, because they
involve a large energy change (it is straightforward to incorporate
the corresponding terms). We also disregarded ripplon
energies $\hbar\omega_q$ compared to $\hbar\omega_{_{\parallel}}$.

A calculation similar to that for crystal lattice defects
\cite{Stoneham75} gives the phase relaxation rate $\Gamma_{\phi}$
for the coupling (\ref{H_iqe}) in the form
\begin{eqnarray}
\label{widthRaman}
\Gamma_{\phi}&&={\pi\over \hbar^2}\sum_{\bf q,q'}|v_{{\bf qq'}1}-v_{{\bf
qq'}2}|^2\nonumber\\
&&\times\bar n(\omega_{\bf q})[\bar n(\omega_{\bf
q'})+1]\delta(\omega_{\bf q}-\omega_{\bf q'}),
\end{eqnarray}
where $\bar n(\omega)=[\exp(\hbar\omega/k_BT)-1]^{-1}$ is the Planck
number. It follows from Eq.~(\ref{widthRaman}) that only thermally
excited ripplons with $\omega_q\lesssim k_BT/\hbar$ contribute to the
rate $\Gamma_{\phi}$.  In what follows we will estimate contributions
to $\Gamma_{\phi}$ from different mechanisms of electron-ripplon
coupling taken separately and will again ignore cross-terms, which
contain products of coupling constants for different mechanisms.

The contribution to the phase relaxation rate $\Gamma_{\phi}^{\rm
(k)}$ from the direct two-ripplon kinematic coupling (\ref{V_2})
has a simple form in the case where the frequencies of ripplons
with $q\lesssim 1/a_{_{\parallel}}$ are small compared to
$k_BT/\hbar$. Then $\Gamma_{\phi}^{\rm (k)}$ is determined
primarily by forward scattering of ripplons off the electron, with
$|{\bf q}-{\bf q}'|\lesssim 1/a_{_{\parallel}}$, but with
$\omega_q=\omega_{q'}\sim k_BT/\hbar \gg \omega_r$. Calculating
the integral over the angle between ${\bf q}$ and ${\bf q}'$ by
the steepest descent method, we obtain
\begin{equation}
\label{phi-k}
\Gamma_{\phi}^{\rm (k)}= {\pi^{1/2}\rho\over
27\sqrt{2}\, a_{_{\parallel}}}\left({k_BT\over
\hbar\sigma}\right)^3 R^2\tilde K_{12}^2,
\end{equation}
where $\tilde K_{12}$ is the difference of the expectation values of
the kinetic energy $p_z^2/2m$ in the states 1 and 2 divided by $R$; we
have $\tilde K_{12}=3/4$ for $E_{\perp}=0$, and $\tilde K_{12}$
decreases with increasing $E_{\perp}$. The numerical value of
$\Gamma_{\phi}^{\rm (k)}$ is very small for low temperatures,
$\Gamma_{\phi}^{\rm (k)}\lesssim 0.7\times 10^2$~s$^{-1}$ for T=10 mK and
$\omega_{_{\parallel}}/2\pi = 20$~GHz.

The contribution from the direct two-ripplon polarization coupling
(\ref{V_2pol}) can be estimated by utilizing the fact that the wave
vectors of thermal ripplons
$q_T=(\rho/\sigma)^{1/3}(k_BT/\hbar)^{2/3}$ are less than $1/r_B$ for
low temperatures. To lowest order in $q_Tr_B$ the polarization
contribution is again given by Eq.~(\ref{phi-k}), but now $\tilde
K_{12}$ is the difference of the expectation values of the potential
energy $\Lambda/2z$ divided by $R$. The corresponding rate is of the
same order as $\Gamma_{\phi}^{\rm (k)}$.

We now estimate the phase relaxation rate due to one-ripplon coupling
(\ref{V}). We note first that the kinematic terms in (\ref{V}) drop
out of the matrix elements $v_{{\bf qq}'j}$, because they do not have
diagonal matrix elements on the functions $|j,0,0\rangle$
($j=1,2$). The terms quadratic in the electric field $E_{\perp}$ drop
out from the difference $v_{{\bf qq}'1} -v_{{\bf qq}'2}$, because they
are independent of the electron state normal to the surface. The major
contribution therefore comes from the polarization one-ripplon
coupling $\propto v_{\rm pol}$ in $\hat V_{\bf q}$ (\ref{V}). We will
denote it as $\Gamma_{\phi}^{\rm (pol)}$.

Polarization terms in the operators $\hat V_{\bf q}$ do not depend on
the in-plane electron coordinate. This makes it possible to calculate
the sum over $\nu, m_{\nu}$ in Eq.~(\ref{qe_coupling}) for $v_{{\bf
qq}'j}$. The calculation is simplified in the case $k_BT\gg \omega_r$
where $q\gg 1/a_{_{\parallel}}$ and ${\bf q\cdot q}'\approx qq'\gg
a_{_{\parallel}}^{-2}$. Then the sum of ${\cal V}^{j\nu m_{\nu}}_{{\bf
q\,q}'}/\hbar \nu\omega_{_{\parallel}}$ can be approximated by
\[2\langle j|\hat V_{-\bf q}|j\rangle\langle j|\hat V_{{\bf q}'}|j\rangle
\exp[-({\bf q}-{\bf
q}')^2a_{_{\parallel}}^2/4]\,(\hbar\omega_{_{\parallel}}a_{_{\parallel}}^2
qq')^{-1}.\]
In this approximation we obtain
\begin{equation}
\label{phi-pol}
\Gamma_{\phi}^{\rm (pol)}\sim {\rho\over
a_{_{\parallel}}}\left({k_BT\over
\hbar\sigma}\right)^3 R^2k_{12}^2.
\end{equation}
Here we assumed that the coefficient $k_{12}=|\langle 1|v_{\rm
pol}(qz)|1\rangle|^2 - |\langle 2|v_{\rm pol}(qz)|2\rangle|^2$ is a
smooth function of $q$ for actual $q\sim q_T$. Its numerical value is
$\sim 0.23$ for $q\approx q_T$ and T=10 mK, it weakly depends on the pressing
field $E_{\perp}$. The phase relaxation rate is $\Gamma_{\phi}^{\rm
(pol)} \sim 10^2$~s$^{-1}$.
The one-ripplon polarization coupling is therefore a major mechanism
of phase relaxation for a confined electron.

The overall ripplon-induced phase relaxation rate appears to be
small. It displays an unusual temperature dependence $T^3$, and
comparatively weakly depends on the in-plane frequency
$\omega_{_{\parallel}}$. We note that it is much less than our
previous estimate \cite{PD99} obtained for the case of in-plane
confinement by a magnetic field.

\section{Ripplon-induced sideband absorption}

An important consequence of coupling to ripplons is the occurrence
of sidebands in the spectrum of microwave absorption by a confined
electron. Sidebands are formed by an electron $|1,0,0\rangle \to
|2,0,0\rangle$ transition accompanied by creation or annihilation
of one or several ripplons. Ripplon sidebands for a confined
electron are similar to phonon sidebands in absorption spectra of
defects in solids \cite{Stoneham75}. They can be understood from
the Franck-Condon picture of an electron transition as happening
for an instantaneous ripplon configuration. Since equilibrium
ripplon positions are different in the ground and excited electron
states, the transition is accompanied by excitation or absorption
of ripplons, and the transition energy differs from its value
$E_2-E_1$ in the absence of coupling to ripplons.

In order to describe the effect it suffices to keep only diagonal in
the relevant electron states $|j,0,0\rangle$ part of the Hamiltonian
of electron-ripplon coupling. One can then apply a standard canonical
transformation which shifts ripplon coordinates so that they are
counted off from their equilibrium values in the ground electron
state. The transformed one-ripplon interaction Hamiltonian
(\ref{H_i1}), (\ref{V}) then takes a Franck-Condon form
\begin{eqnarray}
\label{diagonal}
H_i^{\rm FC}&=&\sum\nolimits_{\bf q}\xi_{\bf
q}\Lambda F(q)|2,0,0\rangle\langle 2,0,0|,\\
F(q)&=&q^2[\langle 2|v_{\rm pol}(qz)|2\rangle -
\langle 1|v_{\rm pol}(qz)|1\rangle]e^{-q^2a^2_{_{\parallel}}/4}.\nonumber
\end{eqnarray}

For weak coupling, of
primary interest are one-ripplon sidebands. Because ripplon
occupation numbers are big for $k_BT \gg \hbar\omega_r$, the
absorption cross-sections for electron transitions accompanied by
absorption and emission of a ripplon are the same. Respectively,
the sidebands are symmetrical as functions of frequency detuning
$\Delta\omega =\omega- (E_2-E_1)/\hbar$ (we have $|\Delta\omega|\sim
\omega_r\ll (E_2-E_1)/\hbar$). The absorption is quadratic in the
electron-ripplon coupling parameters and can be calculated by
perturbation theory in $H_i^{\rm FC}$.
From (\ref{diagonal}) we obtain for the scaled sideband absorption
coefficient $\alpha_{\rm sb}(\omega)$
\begin{eqnarray}
\label{sb_absorption}
&&\alpha_{\rm sb}(\omega)= G_{\rm sb}\bar\alpha_{\rm sb}(\omega),\;
G_{\rm sb}={k_BTR^2r_B^2\rho \over \pi\hbar^2\sigma^2
a_{_{\parallel}}},\nonumber\\
&&\bar \alpha_{\rm sb}= a_{_{\parallel}}\int dq\, q^{-4}
F^2(q)\delta (\Delta\omega\pm\omega_q).
\end{eqnarray}
The full absorption coefficient is given by $\alpha_{\rm sb}$
multiplied by the integral over frequency of the zero-ripplon
absorption coefficient. The halfwidth of the zero-ripplon line
$\Gamma$ is the sum of the decay and dephasing rates calculated
above. In (\ref{sb_absorption}) we assumed that $|\Delta\omega| \gg
\Gamma$; this inequality is well satisfied in the interesting region
$|\Delta\omega|\sim \omega_r$, since from the above estimates
$\omega_r/\Gamma\agt 10^4$.

The intensity of the sideband is determined by the factor $G_{\rm
sb}$. For T=10 mK and $\omega_{_{\parallel}}/2\pi=20$~GHz we have
$G_{\rm sb}\approx 0.1$. The smallness of $G_{\rm sb}$ indicates that
sidebands formed by two- or many-ripplon processes are not important.

\begin{figure}
\includegraphics[width=2.8in]{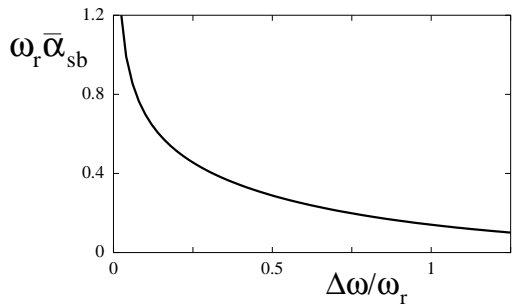}
\caption{Scaled absorption coefficient in the sideband $\bar
\alpha_{\rm sb}$ (\protect\ref{sb_absorption}) vs. frequency detuning
$\Delta\omega = \omega - (E_2-E_1)/\hbar$ for $E_{\perp}=0$ and
$\omega_{_{\parallel}}/2\pi = 20$~GHz. }
\label{fig:sideband}
\end{figure}

The scaled absorption coefficient in the one-ripplon sideband is shown
in Fig.~\ref{fig:sideband}. It monotonically decreases with the
increasing distance $|\Delta\omega|$ from the zero-ripplon line. For
small $|\Delta\omega|$ (but $|\Delta\omega|\gg \Gamma$) we have $\bar
\alpha_{\rm sb} \propto |\Delta\omega|^{-1/3}$. As expected, decay of
the sideband with increasing $|\Delta\omega|$ is much slower than
decay of the Lorentzian tail of the zero-ripplon line $\propto
\Gamma/(\Delta\omega)^2$. For large $|\Delta\omega|/\omega_r$, the
sideband absorption falls off as
$\exp[-(|\Delta\omega|/\omega_r)^{4/3}/2]$, because coupling to
short-wavelength ripplons is exponentially weak. We note that the one-ripplon
sidebands do not display structure, in contrast to sidebands in
electron-phonon systems in solids that reflect singularities in the
phonon density of states.

\subsection{Intensity of the zero-ripplon line}
The integral intensity of the electron absorption spectrum (integral
over frequency of the absorption coefficient) is determined by the
matrix element of the dipolar transition $\propto z_{12}$ and is
independent of the electron-ripplon coupling for
$E_2-E_1\gg\hbar\omega_r$. However, the intensity of the zero-ripplon
line is reduced, because of the sidebands. This reduction is described
by a Debye-Waller type factor (or Pekar-Huang-Reese factor in the
theory of electron-phonon spectra) $\exp(-W)$. The parameter $W$ is
given by the integral of $\alpha_{\rm sb}$ over $\omega$,
\begin{equation}
\label{DW}
W=G_{\rm sb}\bar W,\; \bar W= 2a_{_{\parallel}} \int dq
\,q^{-4}F^2(q).
\end{equation}

\begin{figure}
\includegraphics[width=3.2in]{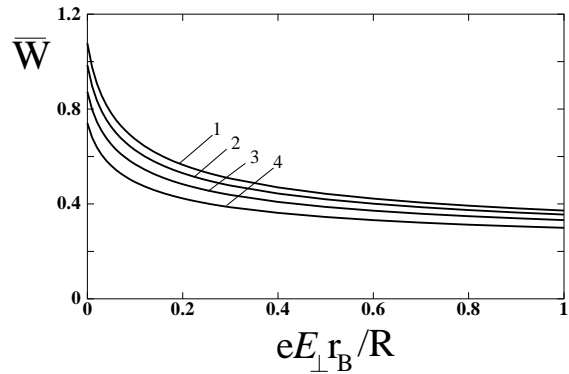}
\caption{Scaling factor $\bar W$ in the expression
(\protect\ref{DW}) for the Debye-Waller exponent $W$. The curves 1
to 4 refer to $r_B/a_{_{\parallel}} = 0.18, 0.25, 0.35, 0.5$
($\omega_{_{\parallel}}/2\pi \approx 10, 20, 39, 79$~GHz, respectively).
}
\label{fig:Debye-Waller}
\end{figure}

The dependence of the scaling factor $\bar W$ on the field $E_{\perp}$
and $\omega_{_{\parallel}}$ is shown in
Fig.~\ref{fig:Debye-Waller}. It is clear from this figure and
Eq.~(\ref{DW}) that $W$ weakly depends on the in-plane electron
frequency $\omega_{_{\parallel}}$ as long as the corresponding ripplon
frequency $\omega_r\ll k_BT/\hbar$. At the same time, $W$ decreases
with the increasing pressing field $E_{\perp}$, because the difference
in the effective radii of the electron states $|1\rangle$ and
$|2\rangle$ decreases, and so does the difference in the ripplon
equilibrium positions in the states $|1\rangle$ and $|2\rangle$.

The calculated value of the Debye-Waller factor for electrostatically
confined electrons $W\approx 0.1-0.05$ for $\omega_{_{\parallel}}/2\pi =
20$~GHz) and $E_{\perp}$ varying from 0 to $\sim 300$~V/cm) is close
to the estimate $W\sim 0.05$ given earlier \cite{PD99} for the case of
in-plane confinement by a magnetic field.  This factor emerges also in
the analysis of the operation of a quantum computer based on trapped
atomic ions \cite{Wineland98}, because optical transitions are
connected to vibrational modes of the ions (the number of such modes
is small, for a small number of ions).

In the context of quantum computing, sideband absorption and the
Debye-Waller reduction of the zero-ripplon absorption strength differ
qualitatively from electron decay and dephasing. The Debye-Waller
mechanism does not affect an electron qubit between quantum
operations. In contrast to dissipative effects, it does not accumulate
between operations. However, it shows that a fraction of electron
transitions may go wrong, as they are accompanied by excitation of
ripplons. The number of such transitions, and therefore the role of
the Debye-Waller factor, can be significantly reduced by using longer
excitation pulses, as we will discuss in a separate publication.

\section{Decay and dephasing from coupling to the electrode}

Coupling to the underlying electrode may also provide an important
mechanism of relaxation of a confined electron (qubit). The
corresponding relaxation parameters can be analyzed in a standard
way. Fluctuations of the electrode potential modulate the inter-level
distance and thus give rise to dephasing. In addition, an electron can
make a transition between the states, with energy being transferred to
an excitation in the electrode (for example, an electron-hole
pair). The analysis is simplified by the fact that the size of the
wave function of the qubit $\sim r_B$ is small compared to the
distance to the electrode $h$. Then the interaction with the electrode
can be described in dipolar approximation,
\begin{equation}
\label{dipolar_coupling}
H_{\rm dip}= -e \,\delta\hat{\cal E}_{\perp} z,
\end{equation}
where $\delta\hat{\cal E}_{\perp}$ is the fluctuating part of the
field on the electron normal to helium surface. This field comes from charge
density fluctuations in the electrode. Here and below we do not
consider effects of fluctuations of the electrode potential on
in-plane motion, which are weak and not important for qubit
dynamics.

Electron relaxation parameters can be expressed in terms of the
correlation function of the fluctuating field
%
\begin{equation}
\label{correlator}
Q(\omega)= \int\nolimits_0^{\infty}dt\,e^{i\omega t}\langle \delta\hat{\cal
E}_{\perp}(t)\delta\hat{\cal E}_{\perp}(0)\rangle.
\end{equation}
We will assume that the function $Q(\omega)$ is smooth in the
frequency ranges of interest for dissipation effects, i.e. for either $\omega
\alt k_BT/\hbar$ or $\omega\sim (E_2-E_1)/\hbar$.

Field-induced time variation of the phase difference of the wave functions
$|1\rangle$ and $|2\rangle$ comes from Stark shift and is equal to
\[\delta\phi_{12}(t)- \delta\phi_{12}(t')=
\hbar^{-1}e(z_{22}-z_{11})\int\nolimits_{t'}^t d\tau\,\delta \hat{\cal
 E}_{\perp}(\tau).\] 
Classical (thermal or quasi-thermal) field fluctuations give rise to
phase diffusion on times that largely exceed the correlation time of
$\delta\hat{\cal E}_{\perp}(t)$, which we assume to be short, $\alt
\hbar/k_BT$.  The coefficient of phase diffusion is equal to the
dephasing rate $\Gamma_{\phi}^{\rm (el)}$. From Eq.~(\ref{correlator})
\begin{equation}
\label{raman_electrode}
\Gamma_{\phi}^{\rm (el)}= e^2(z_{22}-z_{11})^2{\rm Re}~Q(0)/\hbar^2.
\end{equation}
If the noise spectrum Re~$Q(\omega)$ has peaks at low frequencies
$\omega \alt \Gamma_{\phi}^{\rm (el)}$, or the noise $\delta\hat{\cal
E}_{\perp}(t)$ is non-Gaussian, decay of
$\langle\exp[i\delta\phi_{12}(t)]\rangle$ becomes
nonexponential. Although the analysis has to be modified in this case,
it is still  convenient to relate decoherence of electrons on
helium to the fluctuating field $\delta\hat{\cal E}_{\perp}(t)$.

Decay rate of the qubit $\Gamma_{12}^{\rm (el)}$ is
determined by the probability of a field-induced transition $|2\rangle
\to |1\rangle$ between the electron states. This probability is
determined, in turn, by quantum fluctuations of the field $\delta
\hat{\cal E}_{\perp}(t)$ at frequency
$\Omega_{12}=(E_2-E_1)/\hbar$. From Eq.~(\ref{correlator}),
\begin{equation}
\label{decay_electrode}
\Gamma_{12}^{\rm (el)}= e^2|z_{12}|^2{\rm Re}~Q(\Omega_{12})/\hbar^2.
\end{equation}
Here we assumed that decay is due to spontaneous emission only,
i.e. that there are no induced processes with energy transfer
$E_2-E_1$.

To estimate relaxation parameters we will assume that the controlling
electrode is a conducting sphere of a small radius $r_{\rm el}$
submerged at depth $h$ beneath helium surface, as discussed in
Sec.~II. For low frequencies the surface of the sphere is
equipotential.  Then the fluctuating field of the electrode is simply
related to its fluctuating potential $\delta\hat V_{\rm el}$,
$\delta\hat{\cal E}_{\perp} = \delta\hat V_{\rm el}r_{\rm
el}/h^2$. Much of low-frequency noise is due to voltage fluctuations
from an external lead attached to the electrode, which has resistance
${\cal R}_{\rm ext}$ and the temperature $T_{\rm ext}$ that largely
exceeds the helium temperature $T$. The noise can be found from
Nyquist's theorem and gives the dephasing rate
\begin{eqnarray}
\label{spher_electrode_phi}
\Gamma_{\phi}^{\rm (el)}= 2k_BT_{\rm ext}{\cal R}_{\rm
ext}e^2(z_{22}-z_{11})^2 r_{\rm el}^2/\hbar^2h^4.
\end{eqnarray}
For ${\cal R}_{\rm ext}=25$~Ohm, $T_{\rm ext}=1$~K, $r_{\rm
el}=0.1\,\mu$m, $h=0.5\,\mu$m, and $z_{22}-z_{11}=r_B$ we obtain
$\Gamma_{\phi}^{\rm (el)}\approx 1\times 10^4$~s$^{-1}$. This shows
that thermal electrode noise may be a major source of dephasing for a
qubit. The requirement to keep this noise small may be important in
determining the depth by which controlling electrodes can be submerged
below helium surface.

In contrast to low-frequency noise, high-frequency voltage
fluctuations from sources outside the thermostat can be filtered
out. Much of high-frequency quantum fluctuations that affect a qubit
come from the underlying microelectrode itself. They depend on the
interrelation between the electron relaxation time $\tau_{\rm el}$ in
the electrode and $\Omega_{12}^{-1}$. If $\tau_{\rm el}\Omega_{12}\ll
1$, the electrode conductivity does not display dispersion up to
frequencies $\agt \Omega_{12}$; it greatly exceeds $\Omega_{12}$ for
typical $\Omega_{12}$.

An order-of-magnitude estimate of the decay rate $\Gamma_{12}^{\rm
(el)}$ can be made by assuming that the controlling electrode is a
lead attached to a sphere, and this sphere is equipotential
(fluctuations of the total charge in the sphere make a major contribution
to the field $\delta\hat{\cal E}_{\perp}$ for small $r_{\rm
el}/h$). Then from Nyquist's theorem
\begin{equation}
\label{spher_electrode_decay}
\Gamma_{12}^{\rm (el)}= 2(E_2-E_1){\cal R}_{\rm el}e^2|z_{12}|^2r_{\rm
el}^2/\hbar^2h^4.
\end{equation}
where ${\cal R}_{\rm el}$ is the resistance of the lead. If we
estimate it as 0.1~Ohm, then using the same parameters as in the
estimate of $\Gamma_{\phi}^{\rm (el)}$ and setting $E_2-E_1$ equal to
the ``Rydberg'' energy $R$ (\ref{hydrogenic_spectrum}), we obtain
$\Gamma_{12}^{\rm (el)}\sim 5\times 10^2$~s$^{-1}$. 
Even though this estimate is very approximate, it is
clear that the major effect of electrodes on qubit relaxation
is dephasing rather than decay.

\section{Conclusions}

In this paper we have provided a quantitative analysis of the
parameters of qubits based on electrons on helium. We introduced a
simple realistic model of electrodes, which are submerged into helium
in order to localize and control the electrons. This model allowed us
to estimate parameters of the electron energy spectrum and their
dependence on the electrode potential. Control is performed by varying
the field $E_{\perp}$ normal to helium surface. This field changes the
distance between the energy levels of a qubit, which are the ground
and first excited levels of motion normal to the surface, and enables
tuning qubits in resonance with each other and with externally applied
microwave radiation.

The electrode potential  determines not only $E_{\perp}$, but also
the in-plane electron confinement. We found the frequency
$\omega_{_{\parallel}}$ of electron vibrations parallel to helium
surface and related it to the field $E_{\perp}$. Typical frequencies
$\omega_{_{\parallel}}/2\pi$ are of order of a few tens of GHz, and
typical fields are $\sim 100-300$~V/cm. We analyzed both the cases of
one electrode and an electrode array, and investigated the effects of
electrode geometry, including the inter-electrode distance and the
depth by which electrodes are submerged into helium.

We identified relaxation mechanisms, estimated decay rates for a
confined electron, and found their dependence on control
parameters. In contrast to unconfined electrons studied before, decay
is due primarily to electron transitions in which energy is
transferred to two ripplons propagating in opposite directions or to a
bulk phonon propagating nearly normal to the surface.  In both cases
helium excitations with comparatively large wave numbers are involved.
For different coupling mechanisms we have found the dependence of the
decay rate on the parameters of a confined electron. The decay rate
is essentially independent of temperature, for low temperatures.

The overall decay rate is of order $10^4$~s$^{-1}$ for typical
$\omega_{_{\parallel}}$. This estimate is obtained assuming that the
typical wave numbers of excitations into which an electron may scatter
are $\lesssim 10^7$~cm$^{-1}$. We expect that coupling to ripplons and
phonons with much shorter wave lengths is small. Then the decay rate
can be significantly decreased if a magnetic field is applied
perpendicular to helium surface, because such field leads to an
increase in level spacing of in-plane electron excitations, and
therefore more energetic helium excitations have to be involved in
decay.

The major mechanism of dephasing due to coupling to excitations in
helium is scattering of thermal ripplons off an electron. We
calculated the scattering rate and showed that it displays an unusual
$T^3$ temperature dependence.  The major contribution to the dephasing
rate comes from processes which involve virtual transitions between
electron states. The ripplon-induced dephasing rate is $\sim
10^2$~s$^{-1}$ for typical $\omega_{_{\parallel}}$ and $T=10$~mK.

An important mechanism of dephasing is voltage fluctuations of
controlling electrodes. The dephasing rate strongly depends on the
source of these fluctuations and also on the depth by which electrodes
are submerged into helium. An estimate for Johnson noise from a
typical lead connnected to an electrode gives dephasing rate $\sim
10^4$~s$^{-1}$.

We have also analyzed sidebands of electron absorption spectrum
related to electron transitions accompanied by emission or absorption
of a ripplon. We found the Debye-Waller factor which describes the
intensity of the zero-ripplon absorption line and characterizes the
overall probability of exciting a ripplon during an electron
transition. This factor gives fidelity of  qubit excitation
by microwave radiation, which is a major single-gate operation of the
quantum computer based on electrons on helium.

The results provide a quantitative basis for using electrons on helium
as qubits of a quantum computer. The clock frequency of such computer
$\Omega_{\rm QC}$, which is determined by the dipole-dipole
inter-electron interaction, is in the range of $10^7 - 10^8$~Hz even
for inter-electron distances $\approx 1\,\mu$m, and therefore it
largely exceeds both decay and dephasing rates of a confined
electron. Our results suggest ways of further reduction of these
rates. They show how to choose parameters of the system in an optimal
way and also show that there is an extremely broad range where the
parameters can be dynamically controlled, because the inter-level
distance $E_2-E_1 \gg \hbar\omega_{_{\parallel}}\gg \hbar \Omega_{\rm
QC} \gg \hbar\Gamma$.  They also suggest a sequence of steps that have
to be done in order to implement a quantum computer with electrons on
helium in experiment.

\begin{acknowledgments}
We are grateful to A. Dahm, B. Golding, and J. Goodkind for valuable
discussions.  This research was supported by the NSF through grant
No. ITR-0085922.
\end{acknowledgments}

\appendix*
\section{One-ripplon polaronic effect}
Besides relaxation, coupling to ripplons leads also to a polaronic
effect. Because ripplon frequencies are low, the major contribution
comes from processes in which a ripplon is created or annihilated, but
the state of the electron system is not changed. Polaronic shift of
the electron transition frequency is then determined by the diagonal
matrix elements of $H_i^{(1)}$ (\ref{H_i1}) on the wave functions
$|1,0,0\rangle, |2,0,0\rangle$. Keeping only these terms in
$H_i^{(1)}$ corresponds to the adiabatic approximation in which
ripplons have different equilibrium positions depending on the
presence of an electron (one can think of a ``dimple'' made by an
electron on helium surface \cite{Andrei_book}) and on the electron
state.  Of primary interest to us is the state dependence, as it
characterizes the strength of coupling of the electron transition to
ripplons. The corresponding coupling is described by the Franck-Condon
interaction Hamiltonian (\ref{diagonal}).

\begin{figure}
\includegraphics[width=2.8in]{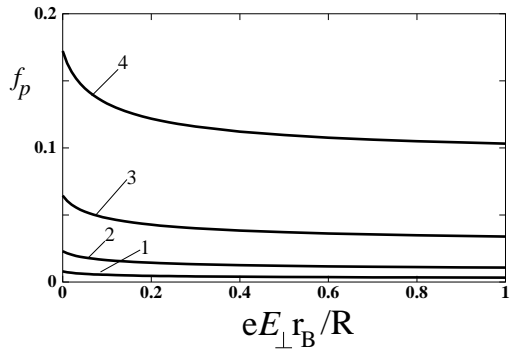}
\hfill

%
\caption{The factor $f_p$ in the Franck-Condon polaronic shift of the
transition frequency of a qubit (\protect\ref{polaron12}) as a
function of the pressing field $E_{\perp}$ for typical values of the
in-plane localization length $a_{_{\parallel}}$. The curves 1 to 4
correspond to $r_B/a_{_{\parallel}}=0.18,0.25,0.35, 0.5$; the
respective values of $\omega_{_{\parallel}}/2\pi$ are $\approx$~10,
20, 39, and 79 ~GHz.}
\label{fig:polaron}
\end{figure}

The Franck-Condon polaronic shift of the transition frequency $1\to 2$
due to the coupling (\ref{diagonal}) is given by a simple perturbation theory,
\begin{eqnarray}
\label{polaron12}
P&=&P_{12}f_p(a_{_{\parallel}}, E_{\perp}),\quad P_{12}=R^2/4
\pi^2\hbar\sigma r_B^2,\\
f_p&=&(4\pi\sigma/\rho)r_B^4
\int\nolimits_0^{\infty}dq\,q^2F^2(q)\omega_q^{-2},\nonumber
\end{eqnarray}
where $f_p$ is a dimensionless factor determined by the matrix
elements of $v_{\rm pol}$ (\ref{V}) on the wave functions of
out-of-plane motion. It depends on the dimensionless parameters
$a_{_{\parallel}}/r_B$ and $eE_{\perp}r_B/R$, and is numerically small
for typical parameter values, see Fig.~\ref{fig:polaron}.
\newline \hspace*{5pt} The numerical value of the factor $P_{12}$ is $P_{12}/2\pi \approx
2.2\times 10^7$~Hz. The energy $\hbar P_{12}$ is much less than the
distance between the electron energy levels. The shift $P$ is also
smaller than the typical frequency $\omega_r$ of ripplons coupled to
the electron (\ref{omega_r}). The inequality $|P| \ll \omega_r$
indicates that the $|1\rangle \to |2\rangle$ transition is weakly
coupled to ripplons.


\begin{thebibliography}{90} 

\bibitem{Loss-DV98} D. Loss and D.P. DiVincenzo, Phys. Rev. A {\bf 57},
120 (1998).

\bibitem{Imamoglu99} A. Imamoglu, D.D. Awschalom, G. Burkard,
D.P. DiVincenzo, D. Loss, M. Sherwin, and A. Small,
Phys. Rev. Lett. {\bf 83}, 4204 (1999).

\bibitem{Si-Ge_spin00} R. Vrijen, E. Yablonovitch, K. Wang, H-W. Jiang,
A. Balandin, V. Roychowdhury, T. Mor and D. DiVincenzo, Phys. Rev. A
{\bf 62}, 012306 (2000)

\bibitem{Kane98} B.E. Kane, Nature {\bf 393}, 133 (1998). 

\bibitem{Yamamoto01} T. D. Ladd, J. R. Goldman, F. Yamaguchi,
Y. Yamamoto, E. Abe, K. M. Itoh, Phys. Rev. Lett. {\bf 89}, 017901 (2002)

\bibitem{Sherwin99} M.S. Sherwin, A. Imamoglu, and Th. Montroy,
Phys. Rev. A {\bf 60}, 3508 (1999).

\bibitem{Steel_Sci00} G. Chen, N.H. Bonadeo, D.G. Steel, D. Gammon,
D.S. Katzer, D. Park, and L.J. Sham, Science {\bf 289}, 1906
(2000).

\bibitem{Piermar01} T.H. Stievater, X.Q. Li, D.G. Steel, D. Gammon,
D.S. Katzer, D. Park, C. Piermarocchi, and L.J. Sham,
Phys. Rev. Lett. {\bf 87}, 133603 (2001).

\bibitem{Averin98} V.D. Averin, Solid State Commun. {\bf 105}, 659
(1998).

\bibitem{Tsai99} Y. Nakamura, Yu.A. Pashkin, and H.S. Tsai, Nature
{\bf 398}, 786 (1999).

\bibitem{Lukens00} J.R. Friedman, V. Patel, W. Chen, S.K. Tolpygo, and
J.E. Lukens, Nature {\bf 406}, 43 (2000).

\bibitem{Orlando00} C.H. van der Wal, A.C.J. ter Haar, F.K. Wilhelm,
R.N. Schouten, C.J.P.M. Harmans, T.P. Orlando, S. Lloyd, and
J.E. Mooij, Science {\bf 290}, 773 (2000).

\bibitem{Han02} Y. Yu, S. Han, Xi Chu, S.-I Chu, and Zh. Wang, Science
{\bf 296}, 889 (2002).

\bibitem{Urbina02} D. Vion, A. Aassime, A. Cottet, P. Joyez,
H. Pothier, C. Urbina, D. Esteve, and M.H. Devoret, cond-mat/0205343.

\bibitem{PD99} P.M. Platzman and M.I. Dykman, Science {\bf 284},
1967 (1999).

\bibitem{DP01} M.I. Dykman and P.M. Platzman, Quantum Information and
Computation {\bf 1},102 (2001)


\bibitem{Lidar_supprt} K.R. Brown, D.A. Lidar, and K.B. Whaley,
Phys. Rev. A {\bf 65}, 012307 (2002)

\bibitem{DeMille02} D. DeMille, Phys. Rev. Lett. {\bf 88}, 067901
(2002).

\bibitem{Andrei_book} {\it ``Two dimensional electron systems on
helium and other cryogenic substrates''}, ed. E.Y. Andrei (Kluwer, 1997).

\bibitem{Shirahama-95} K.~Shirahama, S.~Ito, H.~Suto, and K.~Kono,
J. Low Temp. Phys. {\bf 101}, 439 (1995).

\bibitem{Lea_Fortschr} M.J. Lea, P.G. Frayne, and Yu. Mukharsky,
Fortschr. Phys. {\bf 48} 1109 (2000)


\bibitem{Goodkind01} J.M. Goodkind and S. Pilla,  Quantum
Information and Computation {\bf 1}, 108 (2001).

\bibitem{Dahm_JLT02} A. J. Dahm, J.M. Goodkind, I. Karakurt, and
S. Pilla, J. Low Temp. Phys. {\bf 126}, 709, (2002).

\bibitem{Grimes-76} In the absence of in-plane confinement, electronic
transitions between the states (\protect\ref{hydrogenic_spectrum}) and
the Stark shift of the transition frequency were observed using
microwave spectroscopy by C.C. Grimes, T.R. Brown, M.L. Burns, and
C.L. Zipfel, Phys. Rev. B {\bf 13}, 140 (1976).

\bibitem{jmg_private} J.M. Goodkind, private communication.

\bibitem{Cheng-Cole94} E. Cheng, M.W.~Cole, and M.H.~Cohen,
Phys. Rev. B {\bf 50}, 1136 (1994).

\bibitem{Nieto-00} M.M. Nieto, Phys. Rev. A {\bf 61}, 034901 (2000).

\bibitem{Chuang_book} M.A. Niesen and I.L.~Chuang, {\it Quantum
Computation and Quantum Information} (Cambridge University Press,
Cambridge 2000).



\bibitem{Penanen-Pershan00} Recent experimental data on the $^4$He
liquid-vapor interface were obtained by K.~Penanen, M.~Fukuto,
R.K.~Heilmann, I.F.~Silvera, and P.S.~Pershan, Phys. Rev. B {\bf 62},
9621 (2000). Quantum Monte Carlo simulations for symmetric helium
films were done recently by E.W. Draeger and D.M. Ceperley,
Phys. Rev. Lett. {\bf 89}, 015301 (2002).

\bibitem{Platzman} P.M. Platzman and G. Beni,
Phys. Rev. Lett. {\bf 36}, 626 (1976).

\bibitem{Saitoh} M. Saitoh, J. Phys. Soc. Japan {\bf 42}, 201 (1977).

\bibitem{Monarkha-Shikin82} Yu.P.~Monarkha and V.B.~Shikin,
Fiz. Nizk. Temp. {\bf 8}, 563 (1982)[Sov. J. Low Temp. Phys. {\bf 8},
279 (1982)]




\bibitem{Stoneham75} W. Hayes and A.M. Stoneham, {\it Defects and
defect processes in nonmetallic solids} (Wiley, NY 1985).


\bibitem{Dykman78} M.I. Dykman, Phys. Stat. Sol. (b) {\bf 88}, 463 (1978).

\bibitem{Monarkha78} Yu.P. Monarkha, Fiz. Nizk. Temp. {\bf 4}, 1093
(1978) [Sov. J. Low Temp. Phys. {\bf 4}, 515 (1978)].

\bibitem{Edwards74} D.O. Edwards, J.R.~Eckardt, and F.M. Gasparini,
Phys. Rev. A {\bf 9}, 2070 (1974).

\bibitem{Lastri95} A.~Lastri, F.~Dalfovo, L.~Pitaevskii, and
S.~Stringari, J. Low Temp. Phys. {\bf 98}, 227 (1995).

\bibitem{Wineland98} D.J. Wineland, C.~Monroe, W.M.~Itano,
D.~Leibfried, B.E.~King, and D.M.~Meekhof,
J. Res. Natl. Inst. Stand. Technol. {\bf 103}, 259 (1998).


\end{thebibliography}
\end{document}